\documentclass[12pt]{iopart}%
\usepackage{iopams}
\usepackage{setstack}
\usepackage{graphicx}%
\begin{document}
\title[Scalar Field Cosmology I]{Scalar Field Cosmology I: Asymptotic Freedom and the Initial-Value Problem}

\author{Kerson Huang$^{1,2}$, Hwee-Boon Low$^{2}$, Roh-Suan Tung$^{2}$}

\address{$^{1}$Physics Department, Massachusetts Institute of Technology, Cambridge,
MA, USA 02139}
\address{$^{2}$Institute of Advanced Studies, Nanyang Technological University,
Singapore 639673}
\ead{kerson@mit.edu, rohsuan.tung@ntu.edu.sg}


\begin{abstract}
The purpose of this work is to use a renormalized quantum scalar field to
investigate very early cosmology, in the Planck era immediately following the
big bang. Renormalization effects make the field potential dependent on length
scale, and are important during the big bang era. We use the asymptotically
free Halpern-Huang scalar field, which is derived from renormalization-group
analysis, and solve Einstein's equation with Robertson-Walker metric as an
initial-value problem. The main prediction is that the Hubble parameter
follows a power law: $H\equiv\dot{a}/a\sim t^{-p}$, and the universe expands
at an accelerated rate: $a\sim\exp t^{1-p}$. This gives ``dark energy'', with an
equivalent cosmological constant that decays in time like $t^{-2p}$, which
avoids the ``fine-tuning'' problem. The power law predicts a simple relation for
the galactic redshift. Comparison with data leads to the speculation that the
universe experienced a crossover transition, which was completed about 7
billion years ago.

\end{abstract}
\submitto{\CQG}

\maketitle

\section{Introduction and summary}

According to quantum field theory, the vacuum is not empty and static, but
filled with fluctuating quantum fields. Those of the electromagnetic field,
which fluctuate about zero, can be measured experimentally through the Lamb
shift in the hydrogen spectrum, and the electron's anomalous magnetic moment.
Others, such as the scalar Higgs field of the standard model, fluctuate about
a nonzero vacuum field. Grand unified models call for still more vacuum scalar
fields. These vacuum scalar fields are similar to the Ginsburg-Landau order
parameter in superconductivity, which is a phenomenological way to describe
the condensate of Cooper pairs of the more fundamental BCS theory. Be they
elementary or phenomenological, these vacuum fields behave like classical
fields in many respects. Under certain conditions, however, one must take into
account their quantum nature. In particular, during the big bang, when the
length scale of the universe undergoes rapid change, one must take into
account the effects of renormalization, and this is the focus of the present
investigation. Some of our results have been reported in a previous note [1].

Scalar fields have been used in traditional cosmological theories to explain
``dark energy'' [2], and ``cosmic inflation'' [3]. Dark energy refers to an
accelerating expansion of the universe, which can be reproduced by introducing
a "cosmological constant" in Einstein's equation. This is equivalent to
introducing a static scalar field with constant energy density. The problem is
that the cosmological constant is naturally measured on the Planck scale,
which is some 60 orders of magnitude greater than that fitted to presently
observed data. One would have to ``fine-tune'' it (by 60 orders of magnitude!),
and this has been deemed unpalatable.

The theory of cosmic inflation, designed to explain the presently observed
large-scale uniformity of the universe, postulates that matter was created
while the universe was so small that all matter ``saw'' each other. The universe
then expanded by an enormous order of magnitude (e.g., 27) in an extremely
short time (e.g., 10$^{-26}$s), pushing part of the matter beyond the event
horizon of other parts, but the original density was retained. To implement
this scenario, one introduces a scalar field with spontaneous symmetry
breaking, i.e., having a potential with a minimum located at a nonzero value
of the field. Initially the universe was placed at the ``false vacuum'' of zero
field, and it is supposed to inflate during the time it takes to ``roll down''
the potential towards the true vacuum. It would be desirable to formulate this
scenario in terms of a mathematically consistent initial-value problem.
However, this has not been done so far. As we shall see, the universe does
inflate in our model, but the ``rolling'' was anything but slow.

Most previous works on vacuum scalar fields treat them classically, i.e., with
fixed given potentials. In quantum field theory, however, the potential is
subject to renormalization, and changes with the energy scale. This arises
from the fact that there exist virtual processes with momenta extending all
the way to infinity. The\ high end of the spectrum causes divergences in the
theory, and in any case does not correspond to the true physics. To make the
theory mathematically defined, the spectrum must be cut off at some momentum
$\Lambda,$ and this cutoff is the only scale parameter in a self-contained
field theory. When $\Lambda$ changes, all coupling constants must change in
such a manner as to preserve the theory (i.e., to preserve all the correlation
functions), and this change is called ``renormalization''. Such cutoff
dependence can be ignored when one studies phenomena at a fixed length scale,
such as stellar structure at a particular epoch of the universe; but it is
all-important during the big bang.

The purpose of this work is to study the implications of renormalized quantum
scalar fields in the immediate neighborhood of the big bang. The mathematical
problem is to formulate and solve an initial-value problem based on Einstein's
equation, with suitable idealizations to render the problem tractable. This
basic principle is that there is only one scale in the early cosmos, namely
the ``radius'' $a$ of the universe set by the metric tensor. Thus, we must
identify $a$ with inverse cutoff momentum $\Lambda^{-1}$. For consistency, the
self-interaction potential of the scalar field should be ``asymptotically
free'', i.e., vanish in the limit $a\rightarrow0$.

From renormalization-group (RG) analysis, Halpern and Huang (HH) [4] have
shown that asymptotic freedom determines the potential of the scalar field to
be a Kummer function, a transcendental function that has exponential behavior
for large fields, and this rules out all polynomial potentials, including the
popular $\phi^{4}$ theory. In the present work, we use the HH scalar field as
the source of gravity, in Einstein's equation with Robertson-Walker (RW)
metric. As mentioned earlier, our basic principle is that the inverse radius
of the universe acts as the momentum cutoff of the scalar field theory, i.e.,
$\Lambda=a^{-1}$. This gives rise to a dynamical feedback: the expansion of
the universe is driven by the scalar\ field, whose potential depends on the
radius of the universe.

The main prediction of the model is that the Hubble parameter $H=\dot{a}/a$
behaves like a power $H\sim$ $t^{-p}$ $\left(  0<p<1\right)  $, for large
times, after averaging over small rapid oscillations. The exponent $p$ depends
on model parameters and initial conditions. This indicates ``dark energy'', for
the universe expands with acceleration, according to $a\sim\exp t^{1-p}$. This
behavior corresponds to an equivalent cosmological constant that decays with
time like $t^{-2p}$, and this avoids the usual fine-tuning problem. The origin
of the power law can be traced to a constraint on initial values from the 00
component of Einstein's equation.

Although our model is valid only in a neighborhood of the big bang, it is hard
to resist to compare it with observations from a much later universe. A
partial justification for doing this is that the power-law character may
survive generalizations of the model. In this spirit, we calculate the
relation between luminosity distance $d_{L}$ and red shift $z$ for a light
source, according to the power law. To an extremely good approximation, we
find $d_{L}\left(  z\right)  =z\left(  1+z\right)  d_{0}$, in which the
exponent $p$ enters only through the constant $d_{0}$. Comparison with data on
the galactic redshift, from supernova and gamma-ray burst measurements,
suggest that there was an epoch in which $d_{0}$ had a different value from
the current one, and connecting the two epochs was a crossover transition
completed about 7 billion years ago.

Finally we address the scenario of cosmic inflation, which is inseparable with
matter creation. The question is whether enough matter can be created for
subsequent nucleosynthesis, during the time when the universe was small enough
that all constituents remained within each other's event horizon.

An equally important question has to do with the emergence of the matter
energy scale, which is from the Planck scale by some 18 orders of magnitude.
Physically, the matter scale arises spontaneously, via "dimensional
transmutation" in QCD, and in our model it enters through the coupling
parameter between matter and the scalar field. These two scales must decouple
from each other. How does it happen mathematically in our model?

To explore these questions, we treat matter a perfect fluid coupled to the
scalar field, as detailed in Appendix C. Our studies lead to the opinion that
a completely spatially homogeneous scalar field, real or complex, cannot give
a satisfactory inflation scenario. First, it cannot created enough matter in a
short enough time, and secondly decoupling is not apparent. The model so 
far appears to lack important physical mechanisms in regard to matter creation. 

We are led to investigate a complex scalar field with uniform modulus but
spatially varying phase. This makes the universe a superfluid, and new physics
emerges, namely vorticity and quantum turbulence. We find that these phenomena
can supply the missing mechanisms for matter creation and decoupling. This
development is the subject of paper II of this series [5].

\section{Preliminaries}

We start with Einstein's equation%
\begin{equation}
R_{\mu\nu}-\frac{1}{2}g_{\mu\nu}R=8\pi GT_{\mu\nu}%
\end{equation}
where $g^{\mu\nu}$ is the metric tensor that reduces to the diagonal form
$(-1,1,1,1)$ in flat space-time, $T_{\mu\nu}$ is the energy-momentum tensor of
non-gravitational fields, and $G=6.672\times10^{-11}$ m$^{3}$ kg$^{-1}$
s$^{-2}$ is the gravitational constant. We shall put $4\pi G=1$, thus
measuring everything in Planck units [6]:%
\begin{eqnarray}
\hbox{Planck length}  &  =\left(  \hbar c^{-3}\right)  ^{1/2}\left(  4\pi
G\right)  ^{1/2}=5.73\times10^{-35}\hbox{ m}\nonumber\\
\hbox{Planck time}  &  =\left(  \hbar c^{-5}\right)  ^{1/2}\left(  4\pi
G\right)  ^{1/2}=1.91\times10^{-43}\hbox{ s}\nonumber\\
\hbox{Planck energy}  &  =\left(  \hbar c^{5}\right)  ^{1/2}\left(  4\pi
G\right)  ^{-1/2}=3.44\times10^{18}\hbox{ GeV}%
\end{eqnarray}
Consider a spatially homogeneous universe defined by the Robertson-Walker (RW)
metric, which is specified through the line element%
\begin{equation}
ds^{2}=-dt^{2}+a^{2}(t)\left(  \frac{dr^{2}}{1-kr^{2}}+r^{2}d\theta^{2}%
+r^{2}\sin^{2}\theta d\phi^{2}\right)
\end{equation}
where $t$ is the time, $\{r,\theta,\phi\}$ are dimensionless
spherical coordinates, and $a\left(  t\right)  $ is the length scale. The
curvature parameter is $k=0,\pm1$, where $k=1$ corresponds to a space with
positive curvature, $k=-1$ that with negative curvature, and $k=0$ is the
limiting case of zero curvature. With the RW metric, the $00$ and $ij$
component of Einstein's equation reduce to the following Friedman equations:%
\begin{eqnarray}
\left(  \frac{\dot{a}}{a}\right)  ^{2}+\frac{k}{a^{2}}  &= -\frac{2}{3}%
T_{00}\nonumber\\
\left[  \frac{\ddot{a}}{a}+\left(  \frac{\dot{a}}{a}\right)  ^{2}+\frac
{k}{a^{2}}\right]  g_{ij}  &= -2T_{ij}\hbox{ \ \ \ }(i,j=1,2,3)
\end{eqnarray}
It is customary to introduce the Hubble parameter defined by%
\begin{equation}
H=\frac{\dot{a}}{a}%
\end{equation}
The energy-momentum tensor of a spatially uniform system must have the form%
\begin{eqnarray}
T^{00}  &  =\rho\nonumber\\
T^{ij}  &  =g^{ij}p\nonumber\\
T^{j0}  &  =0
\end{eqnarray}
where $\rho$ defines the energy density, and $p$ the pressure. Energy-momentum
conservation is expressed by $T_{;\mu}^{\mu\nu}=0$, which, with the RW metric,
becomes%
\begin{equation}
\dot{\rho}+\frac{3\dot{a}}{a}\left(  \rho+p\right)  =0
\end{equation}
We can recast the Friedman equations in terms of $H$, and, with inclusion of
the conservation equation, obtain three cosmological equations:%
\begin{eqnarray}
\dot{H}  &  =\frac{k}{a^{2}}-\left(  p+\rho\right) \nonumber\\
H^{2}  &  =-\frac{k}{a^{2}}+\frac{2}{3}\rho\nonumber\\
\dot{\rho}  &  =-3H\left(  \rho+p\right)
\end{eqnarray}
The second equation is a constraint of the form%
\begin{equation}
X\equiv H^{2}+\frac{k}{a^{2}}-\frac{2}{3}\rho=0 \label{constraint}%
\end{equation}
\ The third equation, the conservation law, states $\dot{X}=0$, i.e., the
constraint is a constant of the motion.

As an example, consider Einstein's cosmological constant $\Lambda_{0}$, which
appears in a static energy-momentum tensor of the form (with units restored
for convenience)%
\begin{equation}
T_{0\mu\nu}=-g_{\mu\nu}\left(  \Lambda_{0}/8\pi G\right)
\end{equation}
Corresponding to this, the energy density and pressure are given by%
\begin{eqnarray}
\rho_{0}  &  =\Lambda_{0}/8\pi G\nonumber\\
p_{0}  &  =-\Lambda_{0}/8\pi G
\end{eqnarray}
The conservation equation now states $\dot{\rho}_{0}=0$, which is trivial.
Thus the cosmological equations reduce to%
\begin{eqnarray}
\dot{H}  &  =\frac{k}{a^{2}}\nonumber\\
H^{2}  &  =-\frac{k}{a^{2}}+\frac{2}{3}\rho_{0}%
\end{eqnarray}
The asymptotic solution describes an exponentially expanding universe, with%
\begin{eqnarray}
a\left(  t\right)   &  \sim\exp\left(  H_{\infty}t\right) \nonumber\\
H_{\infty}  &  =\left(  \Lambda_{0}/12\pi G\right)  ^{1/2} \label{generic}%
\end{eqnarray}
Since $a\left(  t\right)  $ is accelerating, we can say that there is ``dark
energy''. However, the ``natural'' value of $H_{\infty}$ should be of order unity
on the Planck scale, whereas the presently observed Hubble parameter is of
order $10^{-60}$. One would have to ``fine tune'' $H_{\infty}$, by sixty orders
of magnitude.

With a dynamical scalar field, the constraint implies $H_{\infty}=0$. This is
illustrated in Appendix A in an exact solution for the massless free scalar
field, in which $\dot{a}/a$ decays according to a power law, which is
equivalent to saying that $H_{\infty}$ decays like a power. The effective
cosmological constant is being "fine-tuned to zero", so to speak. This
"automatic fine-tuning" also happens in our model, to be discussed later.

\section{Halpern-Huang scalar field}

The HH scalar field that we use in this work has an asymptotically free
potential, which is summarized here. Appendix B give a derivation from
renormalization theory.

For generality, consider an $N$-component real scalar field $\phi_{n}\left(
x\right)  $ with $O(N)$ symmetry, with Lagrangian density (with $\hbar=c=1$)%
\begin{equation}
\mathcal{L}_{\hbox{sc}}\left(  x\right)  =-\frac{1}{2}g^{\mu\nu}\sum_{n=1}%
^{N}\partial_{\mu}\phi_{n}\partial_{\nu}\phi_{n}-V\left(  \phi\right)
\label{lagrangian}%
\end{equation}
where $\phi^{2}=\sum_{n=1}^{N}\phi_{n}^{2}$. The high-energy cutoff $\Lambda$
is introduced through a modification of the two-particle propagator at small
distances. (See Appendix B for details.) The form of the modification is not
important here; what is important is that $\Lambda$ is the only intrinsic
scale of the scalar field. All coupling constants $g_{n}$ in the power-series
$V=\sum_{n}g_{n}\phi^{n}$ must scale with appropriate powers of $\Lambda.$ In
4-dimensional space-time we have $g_{n}=\Lambda^{4-n}u_{n}$, where the $u_{n}$
are dimensionless, but depend on $\Lambda$; they undergo "renormalization" in
order to preserve the theory. As $\Lambda$ changes, $\{u_{n}\}$ trace out an
RG trajectory in parameter space. There exist fixed points in this space,
representing scale-invariant systems with $\Lambda=\infty$. A obvious fixed
point is the Gaussian fixed point corresponding to $V\equiv0$, i.e., the
massless free field.

In a universe governed by the RW metric with length scale $a$, we must
identify%
\begin{equation}
\Lambda=\frac{\hbar}{a}%
\end{equation}
where we restore Planck's constant $\hbar$ to remind us of the quantum nature
of the cutoff. The big bang corresponds to $a=0$, or the Gaussian fixed point.
In a consistent theory, therefore, the potential must vanish as $a\rightarrow
0$, or $\Lambda\rightarrow\infty$. In the language of particle physics, the
theory must be ``asymptotically free''. We imagine that at the big bang, the
scalar field was displaced infinitesimally from the Gaussian fixed point onto
some RG trajectory, along some direction in the parameter space. This initial
direction determines the form of $V$. If the trajectory corresponds to
asymptotic freedom, i.e., if the Gaussian fixed point appears as an
ultraviolet fixed point on the trajectory, the potential will grow to engender
a universe. A trajectory that is non-free asymptotically is a critical line on
which all points are equivalent to the fixed point, and the system behaves as
if it had never left the fixed point, with the time development as described
in Appendix A.

All quantities with dimension scale with appropriate powers of $\Lambda.$ The
potential $V$ is of dimensionality (length)$^{-4}$, and we introduce a
dimensionless potential $U$ by writing%
\begin{equation}
V=\Lambda^{4}U
\end{equation}
Under a scale transformation, $U$ changes under renormalization according to%
\begin{equation}
\Lambda\frac{\partial U}{\partial\Lambda}=\beta\left[  U\right]
\end{equation}
where $\beta\left[  U\right]  $ is the ``beta-function'' of the potential.
Near the Gaussian fixed point, where $U\equiv0$, we can make a linear
approximation%
\begin{equation}
\beta\left[  U\right]  \approx-bU
\end{equation}
leading to an eigenvalue equation%
\begin{equation}
\Lambda\frac{dU_{b}}{d\Lambda}=-bU_{b} \label{beta}%
\end{equation}
which defines the eigenpotential $U_{b}$. In the linear approximation, the
most general $U$ is a linear superposition of these eigenpotentials.

From the renormalization-group analysis briefly summarized in Appendix B, one
obtains the solution%

\begin{eqnarray}
U_{b}(z)  &  =c\Lambda^{-b}\left[  M\left(  -2+b/2,N/2,z\right)  -1\right]
\nonumber\\
z  &  =8\pi^{2}\sum_{n}\varphi_{n}^{2} \label{pot}%
\end{eqnarray}
where $M$ is a Kummer function, $c$ is an arbitrary constant, and $\varphi
_{n}\left(  x\right)  $ is a dimensionless field:%
\begin{equation}
\varphi_{n}\left(  x\right)  =\frac{\hbar}{\Lambda}\phi_{n}\left(  x\right)
\end{equation}
Again, we restore units to remind us that the potential depends on $\hbar$.

\eject

The power series and asymptotic behavior of the Kummer function are given by%

\begin{eqnarray}
M(p,q,z)  &  =1+\frac{p}{q}z+\frac{p\left(  p+1\right)  }{q\left(  q+1\right)
}\frac{z^{2}}{2!}+\frac{p\left(  p+1\right)  \left(  p+2\right)  }{q\left(
q+1\right)  \left(  q+2\right)  }\frac{z^{3}}{3!}+\cdots\nonumber\\
M(p,q,z)  &  \approx\Gamma\left(  q\right)  \Gamma^{-1}\left(  p\right)
z^{p-q}\exp z
\end{eqnarray}
Using the derivative formula [7]
\begin{equation}
M^{\prime}\left(  p,q,z\right)  =pq^{-1}M\left(  p+1,q+1,z\right)
\end{equation}
we obtain%
\begin{equation}
U_{b}^{\prime}(z)=-c\Lambda^{-b}N^{-1}\left(  4-b\right)  M\left(
-1+b/2,1+N/2,z\right)
\end{equation}
Asymptotic freedom corresponds to $b>0$ , and spontaneous symmetry breaking
occurs when $b<2$ . Thus we limit ourselves to the range $0<b<2.$

The limiting case $b=2$ corresponds to the massive free field, which is
asymptotically free but does not exhibit spontaneous symmetry breaking, i.e.,
it does not maintain a vacuum field. The limiting case $b=0$ corresponds to
the $\phi^{4}$ theory, which exhibits spontaneous symmetry breaking, but is
not asymptotically free. In our linear approximation (\ref{beta}), $b=0$
corresponds to $\Lambda\partial U/\partial\Lambda=0$, which indicates
neutrality. However, the beta-function to second order gives [8]%
\begin{equation}
\Lambda\frac{\partial U}{\partial\Lambda}=\frac{3}{16\pi^{2}}U^{2}\hbox{
\ \ \ }\left(  \hbox{for }b=0\hbox{, or }\phi^{4}\hbox{ theory}\right)
\end{equation}
which shows it increases as $\Lambda$ increases, and is thus asymptotically non-free.

We emphasize that the HH eigenpotential is derived (a) in flat space-time, (b)
in the neighborhood of the Gaussian fixed point, where $U\equiv0$. Corrections
due to space-time curvature and nonlinearity in $U$ have not been calculated;
but the present approximation should be good in a neighborhood of the big bang.

\section{Cosmological equations}

The canonical Lagrangian (\ref{lagrangian}) of the scalar field gives the
following equation of motion and components of the energy-momentum tensor :
\begin{eqnarray}
\ddot{\phi}_{n} &  =-3H\dot{\phi}_{n}-\frac{\partial V}{\partial\phi_{n}%
}\nonumber\\
\rho_{{canon}} &  =\frac{1}{2}\sum_{n=1}^{N}\dot{\phi}_{n}^{2}%
+V\nonumber\\
p_{{canon}} &  =\frac{1}{2}\sum_{n=1}^{N}\dot{\phi}_{n}^{2}-V\hbox{ }%
\end{eqnarray}
The constraint equation (\ref{constraint}) now reads%
\begin{equation}
X\equiv H^{2}+\frac{k}{a^{2}}-\frac{1}{3}\sum_{n}\dot{\phi}_{n}^{2}-\frac
{2}{3}V=0\label{con}%
\end{equation}
On general principle, the equations of motion must guarantee $\dot{X}=0,$
since it is known that the Cauchy problem in general relativity exists [9].
However, direct computation using $X$ as given in (\ref{con}) yields $\dot
{X}=-\left(  2/3\right)  \dot{a}\left(  \partial V/\partial a\right)  $, which
is nonzero if the cutoff depends on the time. This defect can be attributed to
the fact that the gravitational cutoff has not been built into the Lagrangian
(\ref{lagrangian}). As remedy, we modify $T^{\mu\nu}$ of the scalar field by
adding a term to the pressure, and take
\begin{eqnarray}
\rho &  =\rho_{{canon}}\nonumber\\
p &  =p_{{canon}}-\frac{a}{3}\frac{\partial V}{\partial a}%
\label{modification}%
\end{eqnarray}
The added term gives rise to a ``trace anomaly'':
\begin{eqnarray}
\hbox{Trace anomaly}= - a \frac{\partial V}{\partial a} = \Lambda \frac{\partial V}{\partial \Lambda} \equiv \beta(V)
\end{eqnarray}
That this equals the beta function agrees with results from renormalization theory [10]. We emphasize that this term was added ``by hand", in a manner similar to the addition of the displacement current in Maxwell equation. The deeper reason why this gives the trace anomaly would be a separate investigation. 

For an eigenpotential $V=a^{-4}U_{b}$ it can shown that%
\begin{equation}
a\frac{\partial V}{\partial a}=(b-4)V+\sum_{n}\phi_{n}\frac{\partial
V}{\partial\phi_{n}}%
\end{equation}
The cosmological equations now become%

\begin{eqnarray}
\dot{H}  &  =\frac{k}{a^{2}}-\sum_{n}\dot{\phi}_{n}^{2}+\frac{1}{3}%
a\frac{\partial V}{\partial a}\nonumber\\
\ddot{\phi}_{n}  &  =-3H\dot{\phi}_{n}-\frac{\partial V}{\partial\phi_{n}%
}\nonumber\\
X  &  \equiv H^{2}+\frac{k}{a^{2}}-\frac{1}{3}\sum_{n}\dot{\phi}_{n}^{2}%
-\frac{2}{3}V=0 \label{cosmo}%
\end{eqnarray}
The first two equations now imply $\dot{X}=0$, and we have a closed set of
self-consistent equations.

We are able to work with a set of classical equations, because we have
neglected quantum fluctuations about the vacuum scalar field. However,
important quantum effects are incorporated through the scale dependence of the
potential $V$ arising from renormalization.

\section{Constraint equation and power law}

The constraint equation in (\ref{cosmo}) requires%
\begin{equation}
H=\left(  \frac{2}{3}V+\frac{1}{3}\sum_{n}\dot{\phi}_{n}^{2}-\frac{k}{a^{2}%
}\right)  ^{1/2} \label{constraint1}%
\end{equation}
That $H$ be real and finite imposes severe restrictions on initial values. In
particular, $a=0$ is ruled out; the initial state cannot be exactly at the big
bang. This poses no problem from a practical point of view, for an initial
universe with radius $a\sim1$ (Planck units) is practically a point.

From a physical point of view, we do not expect the model to be valid in the
immediate neighborhood of the big bang, which would be dominated by quantum
fluctuations. The universe could have been created at very high temperatures,
and rapidly cooled through a phase transition to reach a vacuum with
spontaneous broken symmetry. Or it could have been be created in the broken
state. There is no way to know what actually happened; all we know is that we
start our model at some time after the big bang, but still in the Planck era,
with a vacuum field already present.

Now we turn to the consequence of the constraint. Since $V=a^{-4}U$, it would
vanish rather rapidly in an expanding universe. The same is true of $\phi_{n}%
$, which is proportional to $a^{-1}$ by dimension analysis. Thus, the
constraint (\ref{constraint1}) would make $H\rightarrow0$. \ Given the absence
of relevant scale, we expect $H$ to obey a power law:
\begin{eqnarray}
H &  \sim t^{-p}\nonumber\\
a &  \sim\exp t^{1-p}%
\end{eqnarray}
The argument is far from rigorous, of course, but the result is support by the
exactly solution for the massless free field (Appendix A), and is verified in
numerical solutions to be discussed. The latter show that the power law
emerges after averaging over small high-frequency oscillations.

\section{Numerical solutions}

For numerical solutions, we limit ourselves to the simplest case, a real
scalar field $(N=1)$. A multi-component field would yield qualitatively the
same results for a completely uniform universe. It is convenient to rewrite
the cosmological equations as a set of first-order autonomous equations:%
\begin{eqnarray}
\dot{a}  &  =Ha\nonumber\\
\dot{H}  &  =\frac{k}{a^{2}}-v^{2}+\frac{1}{3}a\frac{\partial V}{\partial
a}\nonumber\\
\dot{\phi}  &  =v\nonumber\\
\dot{v}  &  =-3Hv-\frac{\partial V}{\partial\phi}%
\end{eqnarray}
There are 4 unknown functions of time: $a,H,\phi,v$. The initial values must
be real, and satisfy the constraint%
\begin{equation}
H=\left(  \frac{2}{3}V+\frac{1}{3}\dot{\phi}^{2}-\frac{k}{a^{2}}\right)
^{1/2}%
\end{equation}
Although this relation is preserve by the equations, numerical procedures tend
to violate it, and it is difficult to extend time iterations indefinitely. As
an exploratory investigation, we have not looked into algorithm improvement.

For completeness, we restate the HH potential $V$, which is generally a linear
superposition of eigenpotentials $V_{b}$:%
\begin{eqnarray}
V_{b}\left(  \phi\right)   &  =a^{-4}U_{b}(z)\nonumber\\
U_{b}(z)  &  =ca^{b}\left[  M\left(  -2+b/2,1/2,z\right)  -1\right]
\nonumber\\
z  &  =8\pi^{2}a^{2}\phi^{2}%
\end{eqnarray}
where $M$ is the Kummer function. Some useful formulas are%

\begin{eqnarray}
a\frac{\partial V_{b}}{\partial a}  &  =(b-4)V_{b}+\phi\frac{\partial V_{b}%
}{\partial\phi}\nonumber\\
\frac{\partial V_{b}}{\partial\phi}  &  =16\pi^{2}a^{-2}\phi U_{b}^{\prime
}\nonumber\\
U_{b}^{\prime}(z)  &  =-c\left(  4-b\right)  a^{b}M\left(
-1+b/2,3/2,z\right)
\end{eqnarray}
The model parameters are%

\begin{equation}%
\begin{array}
[c]{ll}%
\hbox{Curvature:} & k=1,0,-1\\
\hbox{Eigenvalue:} & 0<b<2\\
\hbox{Potential strength:} & c
\end{array}
\end{equation}
A pair of values $\{b,c\}$ should be specified for each eigenpotential in $V$.
The $c$'s should be real numbers of either sign, such that $V$ be positive for
large $\phi$, and have a lowest minimum at $\phi\neq0.$

First we use an eigenpotential with $b=1$, which is shown in Figure 1 at $a=1$.
As the universe expands, it will increase uniformly by a factor $a(t)$. This
property is a linear approximation that holds for sufficiently small $a(t)$.
Figure 2 shows numerical results for this potential, for curvature parameter
$k=0.$ We see that $H\left(  t\right)  $ oscillates about an average behavior
consilient with a power law $H\sim t^{-p}$, with $p=0.65$. The main source of
uncertainty in $p$ arises from the limitation on time iterations, due to
numerical violation of the constraint. Numerical results for $p$ from a number
of runs are tabulated in Table 1.%

\begin{figure}
[ptb]
\begin{center}
\includegraphics[
width=3in
]%
{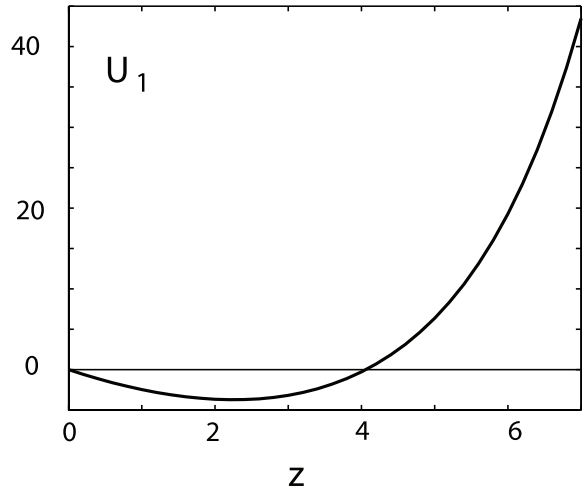}%
\caption{The Halpern-Huang eigenpotential $U_{1}(z),$ with $z=8\pi^{2}\left(
a\phi\right)  ^{2}$, where $\phi$ is a real scalar field, and $a$ is the
Robertson-Walker length scale. The potential increases exponentially for large
$z$.}%
\end{center}
\end{figure}

\begin{figure}
[ptb]
\begin{center}
\includegraphics[
width=5.5in
]%
{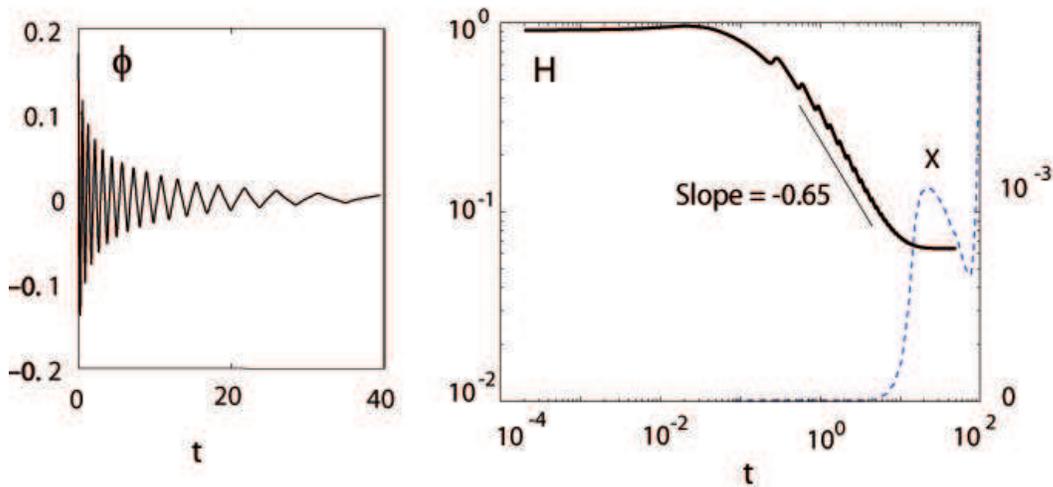}%
\caption{Results from solving the initial-value problem with the potential
$U_{1}$ of Fig.1. The Hubble parameter $H$ follows a power law $t^{-p}$\ after
averaging over small oscillations. The flat tail is spurious, arising from
numerical instability. The scalar field $\phi\ $oscillates with large
amplitudes, sampling the exponential region of the potential. The behavior is
quite different from the ``slow-roll" of conventional inflationary theories. Dotted line shows $X$ of the constraint equation $X=0$ in (31). Numerical violation of the constraint causes the spurious flattening of the curve for H. Note plot for $X$ is a semilog plot, with y-axis label on the right.
}%
\end{center}
\end{figure}

Next we consider a superposition of two eigenpotentials:%
\begin{eqnarray}
U\left(  z\right)   &  =c_{1}U_{b_{1}}\left(  z\right)  -c_{2}U_{b_{2}}\left(
z\right) \nonumber\\
b_{1}  &  =1.6\hbox{, \ \ \ \ \ }c_{1}=0.1\\
b_{2}  &  =0.4\hbox{, \ \ \ \ \ }c_{2}=5.0\nonumber
\end{eqnarray}
The locations $\pm z_{\hbox{min}}$ and the depth $U_{\hbox{min}}$ of the
minima are functions of $a,$ and are plotted in Figure 3. Because of the large
ratio $c_{2}/c_{1}=50,$ $U_{\hbox{min}}$ suddenly jumps at a near-critical
value $a_{\hbox{c}}\approx5$. For $a<a_{\hbox{c}}$, the minima of the can be
approximated by two symmetrically placed $\delta$-functions; the scalar field
becomes trapped at values $\pm\phi_{1}$ corresponding to the minima, and the
model approaches the Ising spin model. Results of numerical solutions are
shown in Figure 4, with curvature parameter is $k=0,$ and the initials
conditions are $a_{0}=1,\phi_{0}=0,\dot{\phi}_{0}=0.1.$

Figure 2 and Figure 4 show that the scalar field oscillates during cosmic expansion,
contrary to the ``slow-roll'' picture of inflation. Closer examination show that
the oscillation amplitudes are so large as to sample the exponential region of
the potential wall. That is, the distinctive part of the HH potential, which
makes it asymptotically free, plays an important role in the expansion of the universe.

\section{Comparison with observations}

Our model is valid only in the Planck era, and does not contain matter apart
from the vacuum scalar field. We shall nevertheless compare the model with
present observations, assuming that the power law $H\left(  t\right)  \sim
h_{0}t^{-p}$ will persist in the real universe. The index $p$ depends on model
parameters, which might change with conditions in the universe such as the
temperature. For our analysis, however, we take $p$ to be a constant. All
quantities are measured in Planck units, unless otherwise specified.

The age of the universe $t_{0}$ and the present value $H_{\hbox{now}}=$
$H\left(  t_{0}\right)  $ are taken to be%
\begin{eqnarray}
t_{0}  &  =1.5\times10^{10}\hbox{ yrs}\approx10^{60}\nonumber\\
H_{{now}}  &  =t_{0}^{-1}%
\end{eqnarray}
The initial value, defined at $t=1,$ is given by%
\begin{equation}
H_{{initial}}=h_{0}\left(  1.65\times10^{50}\right)  ^{-(1-p)}%
\end{equation}
If we put $H_{\hbox{initial}}=1$ as a natural value, then $h_{0}$ gives the
fine-tune factor, which are tabulated Table 1.%

\begin{table}[tbp] \centering
\begin{tabular}
[c]{|l|l|l|l|l|l|l|l|}\hline
$k$ & $b$ & $c$ & $a_{0}$ & $\phi_{0}$ & $\dot{\phi}_{0}$ & $H_{0}$ &
$p$\\\hline
-1 & 1 & 0.1 & 1.00 & 0.01 & 0.1 & 1.00 & 0.81\\
0 & 1 & 0.1 & 1.85 & 0.17 & 0.2 & 0.91 & 0.65\\
1 & 1 & 0.1 & 1.85 & 0.19 & 0.2 & 1.70 & 0.15\\\hline
\end{tabular}
\caption{Computation data:
k = curvature; b,c = potential parameters; others = initial data; p = output exponent. }\label{TableKey copy(2)}%
\end{table}%

The radius of the universe expands according to%
\[
a\left(  t\right)  =a_{0}\exp\frac{h_{0}t^{1-p}}{1-p}%
\]
The present radius is $a(1):$%
\begin{equation}
a_{{now}}=a_{0}\exp\frac{1}{1-p}%
\end{equation}
Some values are tabulated in Table 2.%

\begin{figure}
[ptb]
\begin{center}
\includegraphics[
width=6.6in
]%
{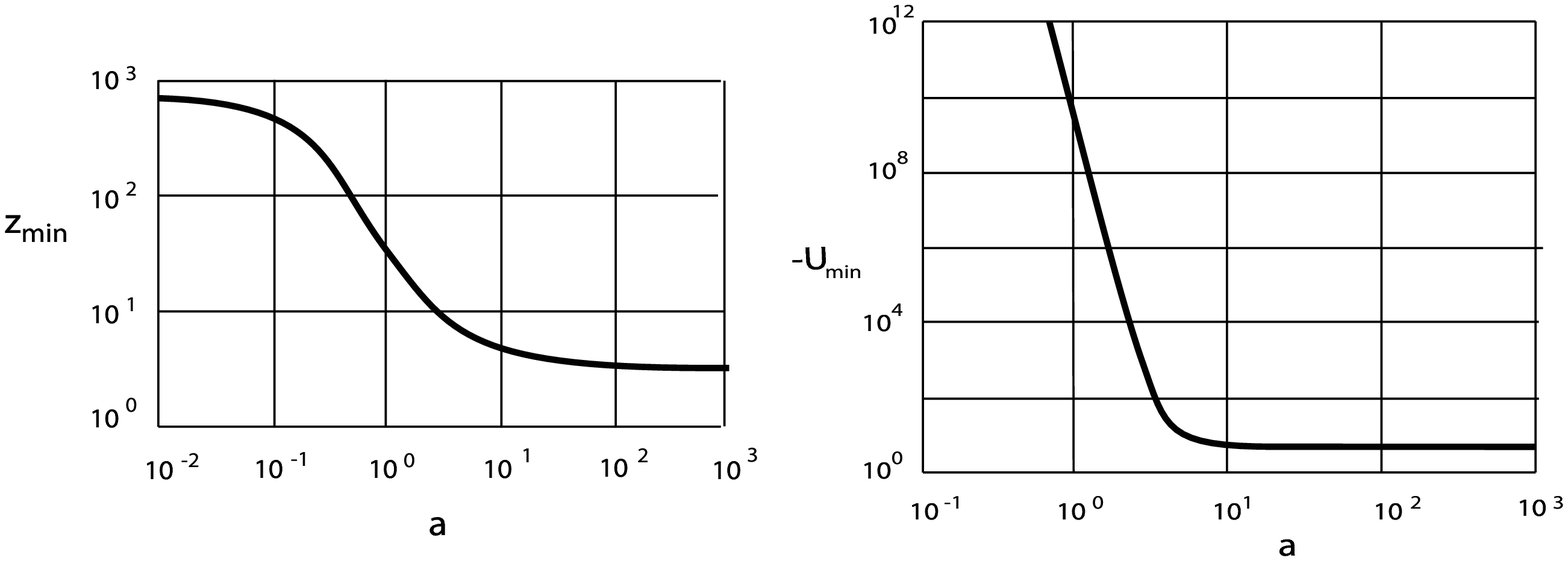}%
\caption{The superposition of two eigenpotentials with a ratio of 50 in relative
strength produces a potential with two symmetrically placed minimum that
approach delta functions in the limit $a\rightarrow0$. The scalar field
becomes trapped in these minima, and the field theory approaches a spin Ising
model. Here, the location of the minima $\pm z_{\hbox{min}}$ and potential
depth $U_{\hbox{min}}$ are plotted as functions of $a$.}%
\end{center}
\end{figure}

\begin{figure}
[ptb]
\begin{center}
\includegraphics[
width=6.5in
]%
{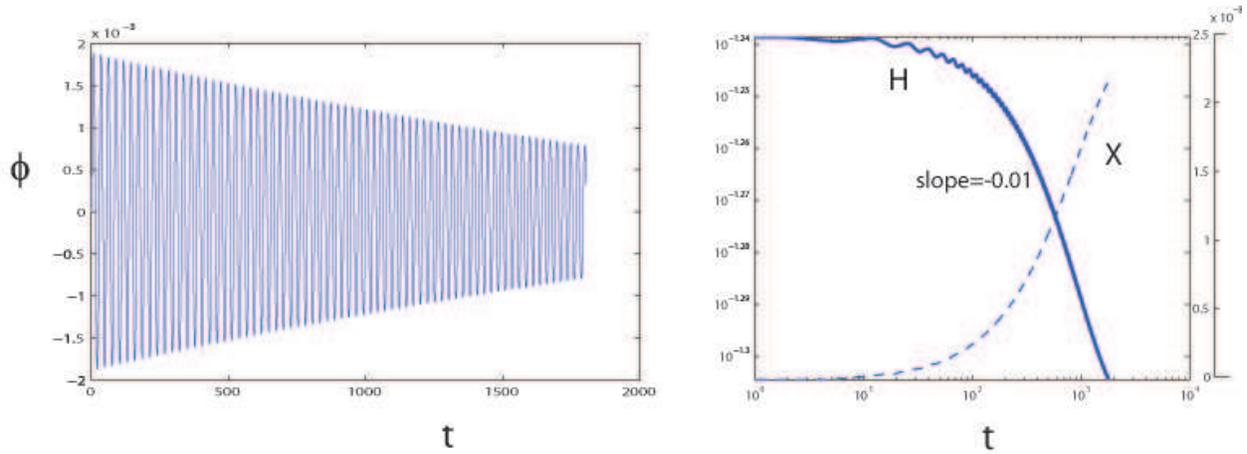}%
\caption{Results from solving the initial-value problem with superposition of
eigenpotentials depicted in Figure 3.}%
\end{center}
\end{figure}

Under the assumption that $p$ is constant, its most reasonable value would lie
in the range $0.99<p<1$.

We now turn to the galactic redshift. The relation between the luminosity
distance $d_{L}$ of the source and the redshift parameter $z$ is implicitly
given by the following relations [11]:%

\begin{eqnarray}
z  &  =\frac{a\left(  t_{0}\right)  }{a\left(  t_{1}\right)  }-1\nonumber\\
f\left(  r_{1}\right)   &  =\int_{t_{1}}^{t_{0}}\frac{dt}{a\left(  t\right)
}\nonumber\\
d_{L}  &  =\frac{r_{1}a^{2}\left(  t_{0}\right)  }{a\left(  t_{1}\right)
}=r_{1}a\left(  t_{0}\right)  \left(  1+z\right)  \label{time1}%
\end{eqnarray}
where $t_{0}$ the the time of detection, at the origin of the coordinate
system, of light emitted at time $t_{1}<t_{0},$ by a source located at
co-moving coordinate $r_{1}$. The function $f$ is defined by%
\begin{equation}
f\left(  r_{1}\right)  \equiv\int_{0}^{r_{1}}\frac{dr}{\sqrt{1-kr^{2}}%
}=\left\{
\begin{array}
[c]{cc}%
\sin^{-1}r_{1} & \left(  k=1\right) \\
r_{1} & \left(  k=0\right) \\
\sinh^{-1}r_{1} & \left(  k=-1\right)
\end{array}
\right.
\end{equation}
Using the first two equations, we can expressed $r_{1}$ and $t_{1}$ in terms
of $t_{0}$ and $z$, and then obtain $d_{L}\left(  z\right)  $ from the third equation.

\begin{table}[tbp] \centering
\begin{tabular}
[c]{|l|l|}\hline
$p$ & $h_{0}$\\\hline
0.5 & $1.25\times10^{25}$\\\hline
0.85 & $3\times10^{7}$\\\hline
0.95 & $300$\\\hline
0.99 & $3$\\\hline
\end{tabular}
\caption{Fine-tune factor for Hubble's parameter}\label{TableKey}%
\end{table}%

\begin{table}[tbp] \centering
\begin{tabular}
[c]{|l|l|}\hline
$p$ & $a_{\hbox{now}}/a_{0}$\\\hline
0.5 & $7.4$\\\hline
0.85 & $786$\\\hline
0.95 & $5\times10^{8}$\\\hline
0.99 & $3\times10^{43}$\\\hline
\end{tabular}
\caption{Present radius of universe}\label{TableKey copy(1)}%
\end{table}%

In our model, $a\left(  t\right)  =a_{0}\exp\left(  \xi t^{1-p}\right)
,$where $\xi=h_{0}\left(  1-p\right)  ^{-1}$. Define an effective time
$\tau=\xi t^{1-p}$. For $0<p<1$, the second equation in (\ref{time1}) can be
rewritten as%
\begin{equation}
f\left(  r_{1}\right)  =K_{0}\int_{\tau_{1}}^{\tau_{0}}d\tau\tau^{p/\left(
1-p\right)  }\exp\left(  -\tau\right)  \label{time2}%
\end{equation}
where $K_{0}=\left[  \left(  1-p\right)  a_{0}\right]  ^{-1}\xi^{-1/\left(
1-p\right)  },$ and%
\begin{eqnarray}
\tau_{0}  &  =\xi t_{0}^{1-p}\nonumber\\
\tau_{1}  &  =\tau_{0}-\ln\left(  z+1\right)  \label{time3}%
\end{eqnarray}
Since $t_{0}\approx10^{60}$, we can assume $\tau_{0}>>1$, and obtain to a good
approximation $f\left(  r_{1}\right)  \approx K_{1}z,$where $K_{1}=K_{0}%
\tau_{0}^{p/\left(  1-p\right)  }\exp\left(  -\tau_{0}\right)  $. Since
$K_{0}$ is extremely small, this gives $r_{1}=z$ to a very good approximation,
and thus%
\begin{equation}
d_{L}=K_{1}a_{0}z\left(  1+z\right)
\end{equation}
We rewrite this as%
\begin{equation}
\frac{d_{L}}{z}=d_{0}\eta\left(  1+z\right)  \label{dl/z}%
\end{equation}
where $d_{0}=c/H_{\hbox{now}}=4283$ Mpc, corresponding to the choice
$H_{\hbox{now}}=70$ km s$^{-1}$Mpc$^{-1}.$

\begin{figure}
[ptb]
\begin{center}
\includegraphics[
width=5in
]%
{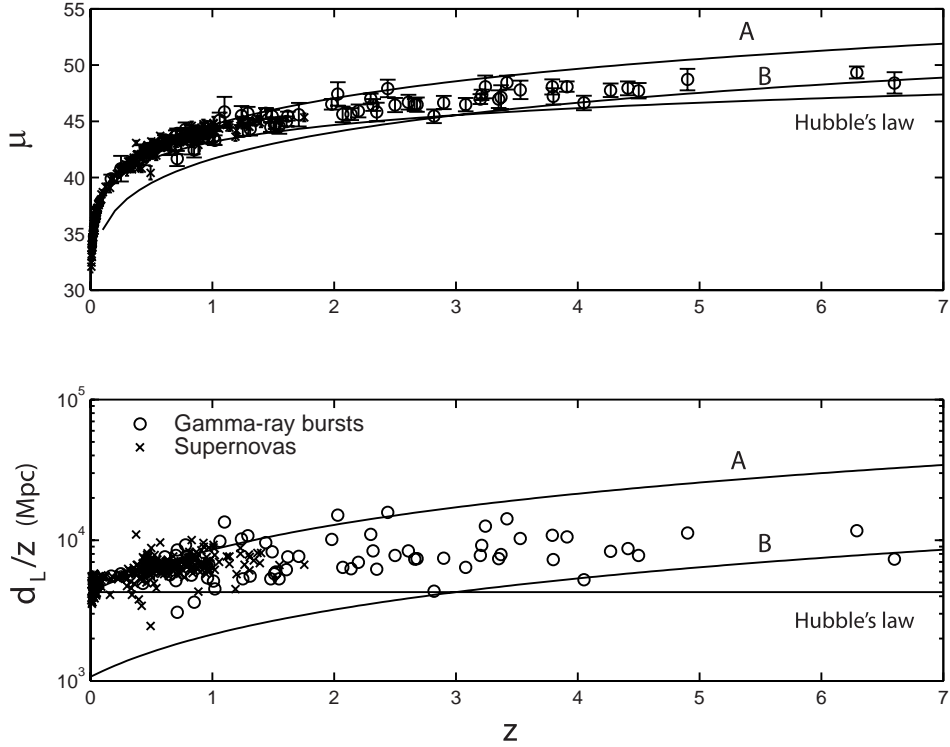}%
\caption{Comparison between model prediction of the galactic redshift with
observational data. Upper panel shows a conventional plot. Lower panel show a
log-log plot of $d_{L}/z$ vs $z$, where $d_{L}$ is the luminosity distance and
$z$ is the redshift parameter. The two theoretical curves, labeled A and B,
correspond to different values of the exponent $p$, which depends on
parameters in the scalar potential, and initial conditions. See text for
fuller explanation.}%
\end{center}
\end{figure}

Figure 5 shows comparison with data from observations and supernovas [12] and
gamma-ray bursts [13]. The upper panel shows the parameter $\mu$ used in
conventional data analysis:%
\begin{equation}
\mu=5\log\left(  \frac{d_{L}}{\hbox{Mpc}}\right)  +25
\end{equation}
plotted as a function of z. The lower panel shows a semilog plot of $d_{L}/z$
vs. $z$. Lines corresponding to Hubble's law (no dark energy) are shown. The
$p$-dependence affects only the vertical displacement but not the shape of the
model curves. Curve A corresponds to (\ref{dl/z}) with $\eta=1,$ and curve B
with $\eta=1/4.$ Curve A fits the data for $z<1$, while curve B could
represent the situation in a large-$z$ regime beyond present measurements.

The power-law model allows only for variations in $d_{0}$, which may come from
variations in the exponent $p$, caused by conditions such as the temperature.
This leads us to speculate that the universe may have had gone through a broad
phase transition, or crossover, connecting two situations corresponding
respectively to the curves A and B. The transition was completed around $z=1$.

The relation between the emission time and the red shift can be obtained from
(\ref{time3}):
\begin{equation}
\frac{t_{1}}{t_{0}}=\left[  1-\left(  1-p\right)  \ln\left(  z+1\right)
\right]  ^{1/\left(  1-p\right)  }%
\end{equation}
For $p\approx1$, we put $p=1-\epsilon$ and obtain%
\begin{equation}
\frac{t_{1}}{t_{0}}\approx\left[  1-\epsilon\ln\left(  z+1\right)  \right]
^{1/\epsilon}\underset{\epsilon\rightarrow0}{\longrightarrow}\left(
z+1\right)  ^{-1}%
\end{equation}
Assuming this relation, we judge that the transition was completed at
$t_{1}/t_{0}\approx0.5$, or about than 7 billion years ago.

\section{Cosmic inflation and decoupling}

The problem of cosmic inflation is inseparable from that of matter creation,
which has not been taken into account in our model so far. Most of the matter
in the universe should have been created by the end of the inflation era, in
order that the memory of the original density be imprinted.

An equally important problem relates to energy scales. Our equations so far
contains only one scale, the Planck scale. With matter creation, there emerges
the scale of nuclear interactions, which is smaller than the Planck scale by
some 18 orders of magnitude. Physically speaking, the matter scale emerges in
QCD spontaneously through ``dimensional transmutation'' [14]. In our model, we
would introduce it through the coupling parameter between the scalar field and
matter. These two scales must eventually decouple from each other. That is,
the cosmological equations should break up into two sets, one governing the
expansion, the other galactic evolution, and in each set the information about
the other set occurs only through lumped constants. What is the mechanism for
this decoupling?

To address these questions, we model matter as a perfect fluid coupled to the
scalar field, and obtain a set of cosmological equations that, again,
represent an initial-value problem. These are derived in Appendix C. Numerical
studies of these equations, both for a real scalar field and a complex scalar
field, lead us to the conclusion that a completely uniform scalar field, even
with more than one components, cannot create sufficient matter to satisfy the
inflation scenario. More important, it cannot exhibit the decoupling desired.

We are led to an attempt to relax complete uniformity, within the dictate of
the RW metric. It is natural to consider a complex scalar field with uniform
modulus, but spatially varying phase. The phase variation gives rise to
superfluid velocity, with the attendant vortex dynamics. The universe then
becomes a superfluid with vortex dynamics. New physics emerges, namely the
growth and decay of a vortex tangle that fills the universe, signifying
quantum turbulence. We find that this provides a framework for matter
creation, and the decoupling of scales.

In the extension of our model, the demise of quantum turbulence will signifies
the end of the inflation era, as well as the validity of our model, for
density fluctuations would become important. The standard hot big bang theory
will then take over, with one addition: the universe remains a superfluid with
vorticity. The latter will offer explanations to post-inflation phenomena such
as galactic voids, galactic jets, and the dark mass. We will present this
development in detail in paper II of this series [5].

\section{Critical comments}

We address some open issues in this investigation.

\begin{enumerate}

\item
The HH potential is derived in flat space-time, to lowest order in the
scale parameter $a=\Lambda ^{-1}$ . To calculate corrections, one can use
Polchinski's equation (B.6) in Appendix B. Preliminary calculations indicate
no qualitative change to the next power in $a,$ for the Robertson-Walker
metric with $k=0$.

\item
One can use a linear combination of the HH eigenpotentials, and this
gives greater freedom in model building. The potential corresponds to the
direction along which the RG trajectory emerges from the Gaussian fixed
point in parameter space, at the big bang. One can only say that it is the
one that spawns our universe, that is, the parameters are phenomenological.
One should make a general survey of possible models. In this respect, one
faces the well-known but unsolved computational problem of how to maintain
the integrity of a conservation law under indefinite time iterations.

\item
Because of the problem mentioned above, the preliminary calculations
reported in this paper are subject to numerical uncertainty. However, they
do point to a common feature, the power law. This is due to fact that the
constraint limits the initial data to a special subset. Outside of this
subset, one gets an exponential law. This behavior is clearly illustrated in
the exact solution of the massless scalar field in Appendix A.

\item
We assume that, at the initial time when our model takes hold, which is
shortly after, but not exactly at, the big bang, the vacuum field exists,
and we can ignore any other form of matter present, for the purpose of
studying the expansion of the universe. It could happen that the big bang
delivered a very hot universe, which cooled down to this state in a very
short time, through a phase transition. But it could also happen that the
big bang delivered a cold universe. All we care is that the mathematical
model starts with the conditions stated. Einstein's cosmological constant,
though not introduced in the cosmological equations, arises effectively
through the initial conditions on the vacuum field.

\item
Our model is semi-classical, in the sense that we ignore quantum
fluctuations about the vacuum scalar field. Some quantum effects are
included, namely renormalization, and the relation $a=\Lambda ^{-1}.$ We
wish to expand on the latter point. This relation can be immediately
implemented in the Robertson-Walker metric, but how would one do it in a
general metric? One would have to go back to the joint action of gravitation
field $g^{\mu \nu }$ and scalar field $\phi $. The cutoff $\Lambda $ enters
through the kinetic term $\phi K^{-1}\phi $ in the Lagrangian density, where 
$K^{-1}\rightarrow $ $g^{\mu \nu }\partial _{\mu }\partial _{\nu }$ at
distances large compared to a radius of order $\Lambda ^{-1}$. The scale
parameter $a$ is contained in $g^{\mu \nu }$. Thus, $a=\Lambda ^{-1}$ says
that gravity cuts off virtual processes in the scalar field. Given an
explicit form of $K^{-1}$, the classical variational principle will yield
coupled Einstein-scalar equations. The preceding procedure is
semi-classical, because $\phi $ is the vacuum field without quantum
fluctuations. In a complete quantum field theory, one must insert the
classical action into the Feynman path integral for transition amplitudes,
and functionally integrate over all possible $g^{\mu \nu }$ and $\phi $.
Needless to say, one would encounter the difficulties of quantum gravity.
For our purpose, fortunately, we can implement $a=\Lambda ^{-1}$ without
getting in too deeply.

\item
We added a term $\left( a/3\right) \partial V/\partial a$ to the
canonical pressure in (28), in order to preserve the constraint $X=0.$ This
was done strictly ``by hand'', in a spirit similar to Maxwell's introduction
of the displacement current. It is gratifying that this leads to the trace
anomaly, but we don't really understand why. The trace anomaly is a purely
quantum effect, arising from a rather subtle property of the Feynman path
integral, namely that the measure in the functional integration acquires a
certain phase factor under a scale transformation. A deeper understanding of
this term may give insight into quantum gravity.

\end{enumerate}

\appendix

\section{The massless free field}

The cosmological equations with a real massless scalar field, corresponding to
$V\equiv0$, are%
\begin{eqnarray}
\dot{a}  &  =Ha\nonumber\\
\dot{H}  &  =\frac{k}{a^{2}}-\dot{\phi}^{2}\nonumber\\
\ddot{\phi}  &  =-3H\dot{\phi}\nonumber\\
X  &  \equiv H^{2}-\frac{1}{3}\dot{\phi}^{2}+\frac{k}{a^{2}}=0
\end{eqnarray}
They describe what happens if the scalar field remains at the Gaussian fixed
point. The last equation $X=0$ is the constraint equation, and $X$ is a
constant of the motion.

The third equation can be rewritten in the form $d\ln\left(  \dot{\phi}%
a^{3}\right)  /dt=0$, which gives%
\begin{equation}
\dot{\phi}=c_{0}a^{-3}%
\end{equation}
where $c_{0}$ is an arbitrary constant. The equations then reduce to%
\begin{eqnarray}
\dot{a}  &  =Ha\nonumber\\
\dot{H}  &  =\frac{k}{a^{2}}-\frac{c_{0}^{2}}{a^{6}}\nonumber\\
H^{2}  &  =\frac{c_{1}}{a^{6}}-\frac{k}{a^{2}}%
\end{eqnarray}
where $c_{1}=c_{0}^{2}/3$. Dividing the second equation by the first, and
equating $\dot{H}/\dot{a}=$ $dH/da,$we obtain%
\begin{equation}
HdH=\left(  \frac{k}{a^{3}}-\frac{c_{0}^{2}}{a^{7}}\right)  da
\end{equation}
Integrating both sides gives%
\begin{equation}
H=\pm\sqrt{\frac{c_{1}}{a^{6}}+c_{2}-\frac{k}{a^{2}}}%
\end{equation}
Since $H=\dot{a}/a$, this can be further integrated to yield%
\begin{equation}
t=\pm\int\frac{da}{\sqrt{c_{1}a^{-4}+c_{2}a^{2}-k}} \label{noncon}%
\end{equation}
where $c_{2}$ is an arbitrary constant. The $\pm$ signs reflect the
time-reversal invariance of the equations. We choose the positive sign to
obtain
\begin{equation}
a\left(  t\right)  \underset{t\rightarrow\infty}{\longrightarrow}a_{0}%
\exp\left(  \sqrt{c_{2}}t\right)  \label{nocon1}%
\end{equation}
This is the general solution without constraint, and $c_{2}$ is the equivalent
cosmological constant.

\bigskip The constraint equation can be put in the form%
\begin{equation}
\frac{\dot{a}}{a}=\pm\sqrt{c_{1}a^{-6}-ka^{-2}}%
\end{equation}
which gives%
\begin{equation}
t=\pm\int\frac{da}{\sqrt{c_{1}a^{-4}-k}}%
\end{equation}
Comparison with (\ref{noncon}) shows%

\begin{equation}
c_{2}=0
\end{equation}
Thus, (\ref{nocon1}) is incorrect; the constraint ``fine-tunes'' the
cosmological constant to zero. The correct solution gives%
\begin{equation}
a(t) = \left\{
\begin{array}
[c]{ll}%
c_{1}^{-1/6}t^{1/3} & \hbox{\ \ }(k=0)\\
\underset{t\rightarrow\infty}{\longrightarrow}c_{1}^{-1/4} & \hbox{\ \ }%
(k=1)\\
\underset{t\rightarrow\infty}{\longrightarrow}t\hbox{ } & \hbox{\ }(k=-1)
\end{array}
\right.
\end{equation}
which corresponds to a power-law%
\begin{equation}
H\underset{t\rightarrow\infty}{\longrightarrow}h_{0}t^{-1}%
\end{equation}

\section{Renormalization and the Halpern-Huang potential}

A distinctive feature of quantum field theory is that the field can propagate
virtually. This is described by the propagator function, which for a free
field has Fourier transform $\Delta\left(  k^{2}\right)  =k^{-2}$. The
high-$k$, or high-energy modes must be cut off, for otherwise the virtual
processes lead to divergences, rendering the quantum theory meaningless. The
cut off energy $\Lambda$ is introduced by "regulating" the propagator:%
\begin{eqnarray}
&  \Delta\left(  k^{2}\right)  =\frac{f(k^{2}/\Lambda^{2})}{k^{2}}\nonumber\\
&  f\left(  z\right)  \underset{z\rightarrow\infty}{\rightarrow}0
\end{eqnarray}
The detailed form of $f(k^{2}/\Lambda^{2})$ is not important. What is
important is that $\Lambda$ is the only scale in the theory. The regulated
propagator in configurational space will be denoted by $K\left(
x,\Lambda\right)  .$

In the formulation of renormalization according to Wilson [15,16], interaction
coupling parameters must change with $\Lambda,$ in such a fashion as to
preserve the theory. This is called ``renormalization''. For a given value of
$\Lambda$, the parameters define an effective theory appropriate to that
energy scale. A reformulation of the Wilson scheme using functional methods
has been given by Polchinski [17].

Interactions that go to zero in the short-distance limit (or infinite-energy
limit) are said to be asymptotically free, an example of which is the gauge
interaction in QCD. In the opposite non-free behavior, the interactions grow
indefinitely with decreasing length scale, and would diverge in the limit.
This is the behavior found in QED and the $\phi^{4}$ scalar field, for which
the short-distance limit can exist only if there is no interaction at all. For
applications in cosmology, we want interactions that vanish at the big bang,
the small-distance limit, which means asymptotically free interactions.

The Halpern-Huang (HH) potential was originally derived [4] by summing
one-loop Feynman graphs. Here we outline an improved derivation due to Periwal
[18], which is based on Polchinski's functional method of renormalization. For
simplicity consider a real scalar field $(N=1).$ The action in $d$-dimensional
Euclidean space-time can be written as
\begin{equation}
S[\phi,\Lambda]=S_{0}[\phi,\Lambda]+S^{\prime}[\phi,\Lambda]\label{kin}%
\end{equation}
where the first term corresponds to the free field, and the second term
represents the interaction. We have%
\begin{equation}
S_{0}\left[  \phi,\Lambda\right]  =\frac{1}{2}\int d^{d}xd^{d}y\,\phi\left(
x\right)  K^{-1}\left(  x-y,\Lambda\right)  \phi\left(  y\right)
\end{equation}
where $K^{-1}\left(  x-y,\Lambda\right)  $ is the inverse of the propagator
$K(x-y,\Lambda)$, in an operator sense. It differs from the Laplacian operator
significantly only in a neighborhood of $\left\vert x-y\right\vert =0$, of
radius $\Lambda^{-1}$. The partition function with external source $J,$ which
generates all correlation functions of the theory, is given by%
\begin{equation}
Z[J,\Lambda]=\mathcal{N}\int D\phi e^{-S[\phi,\Lambda]-(J,\phi)}%
\label{partition}%
\end{equation}
where $\mathcal{N}$ is a normalization constant, which may depend on $\Lambda
$, and $(J,\phi)=\int d^{d}xJ\left(  x\right)  \phi\left(  x\right)  $.

In Wilson's renormalization scheme, modes contributing to the integral in
(\ref{partition}) with momentum higher than $\Lambda$ are \textquotedblleft
integrated out", but not discarded, in order to lower the effective cutoff.
This leads to a change the form of $S^{\prime}$, but the system itself is
unaltered. The interactions are then said to be "renormalized". In a general
sense, renormalization means changing the cutoff $\Lambda$ with simultaneous
change in the form of $S^{\prime}$, so as to leave $Z$ invariant, i.e.,%
\begin{equation}
\frac{dZ[J,\Lambda]}{d\Lambda}=0 \label{condition}%
\end{equation}

\noindent\ This constraint is solved by Polchinski's renormalization
equation\textit{,}which is\textit{ }a functional integro-differential equation
for $S^{\prime}\left[  \phi,\Lambda\right]  $\textit{. }For $J\equiv0,$ it
reads
\begin{equation}
\frac{dS^{\prime}}{d\Lambda}=-\frac{1}{2}\int dxdy\frac{\partial K\left(
x-y,\Lambda\right)  }{\partial\Lambda}\left[  \frac{\delta^{2}S^{\prime}%
}{\delta\phi\left(  x\right)  \delta\phi\left(  y\right)  }-\frac{\delta
S^{\prime}}{\delta\phi\left(  x\right)  }\frac{\delta S^{\prime}}{\delta
\phi\left(  y\right)  }\right]  \label{pol}%
\end{equation}
 Assuming that there are no derivative couplings, we can write
$S^{\prime}$ as the integral of a local potential:%
\begin{eqnarray}
S^{\prime}\left[  \phi,\Lambda\right]   &  =\Lambda^{d}\int d^{d}xU\left(
\varphi\left(  x\right)  ,\Lambda\right) \nonumber\\
\varphi\left(  x\right)   &  =\Lambda^{1-d/2}\phi\left(  x\right)
\end{eqnarray}
where $U$ is a dimensionless function, and $\varphi$ is a dimensionless field.
In the neighborhood of the Gaussian fixed point, where $S^{\prime}=0$, we can
linearize (\ref{pol}) by neglecting the last term, and obtain a linear
differential equation for $U\left(  \varphi,\Lambda\right)  $:%
\begin{equation}
\Lambda\frac{\partial U}{\partial\Lambda}+\frac{\kappa}{2}U^{\prime\prime
}+\left(  1-\frac{d}{2}\right)  \varphi U^{\prime}+Ud=0
\end{equation}
where a prime denote partial derivative with respect to $\varphi$, and
$\kappa=\Lambda^{3-d}\partial K\left(  0,\Lambda\right)  /\partial\Lambda$.
Now we seek eigenpotentials $U_{b}\left(  \varphi,\Lambda\right)  $ with the
property%
\begin{equation}
\Lambda\frac{\partial U_{b}}{\partial\Lambda}=-bU_{b} \label{eigen}%
\end{equation}
In the language of perturbative renormalization theory, the right side is the
linear approximation to the $\beta$-function. Substituting this into the
previous equation, we obtain the differential equation
\begin{equation}
\left[  \frac{\kappa}{2}\frac{d^{2}}{d\varphi^{2}}-\frac{1}{2}\left(
d-2\right)  \varphi\frac{d}{d\varphi}+\left(  d-b\right)  \right]  U_{b}=0
\label{diffequ}%
\end{equation}
 Since this equation does not depend on $\Lambda,\ $the $\Lambda
$-dependence of the potential is contained in a multiplicative factor. In view
of (\ref{eigen}), the factor is $\Lambda^{-b}$.

For $d\neq2,$ (\ref{diffequ}) can be transformed into Kummer's
equation:%
\begin{equation}
\left[  z\frac{d^{2}}{dz^{2}}+\left(  q-z\right)  \frac{d}{dz}-p\right]
U_{b}=0
\end{equation}
where%
\begin{eqnarray}
q  &  =1/2\nonumber\\
p  &  =\frac{b-d}{d-2}\nonumber\\
z  &  =\left(  2\kappa\right)  ^{-1}\left(  d-2\right)  \varphi^{2}%
\end{eqnarray}
The solution is%
\begin{equation}
U_{b}\left(  z\right)  =c\Lambda^{-b}\left[  M\left(  p,q,z\right)  -1\right]
\end{equation}
where $c$ is an arbitrary constant, and $M$ is the Kummer function. We have
subtracted $1$ to make $U_{b}\left(  0\right)  =0$. This is permissible, since
it merely changes the normalization of the partition function. In (\ref{pot}),
the value of $\kappa$ corresponds to a sharp cutoff.

For $d=2$, the solution to (\ref{diffequ}) is sinusoidal, and the theory
reduces to the XY model, or equivalently the so-called sine-Gordon theory [19].

\section{Coupling to perfect fluid}

We discuss how the cosmological equations (\ref{cosmo}) may be generalized to
include coupling to galactic matter modeled as a perfect fluid, whose
energy-momentum tensor is given by [20]%
\begin{equation}
T_{{m}}^{\mu\nu}=-g^{\mu\nu}\rho_{{m}}+\left(  p_{{m}}%
+\rho_{{m}}\right)  U^{\mu}U^{\nu}%
\end{equation}
where $\rho_{{m}}$ is the energy density, and $U^{\mu}$ is a velocity
field$,$ with $g_{\mu\nu}U^{\mu}U^{\nu}=1$. For a spatially uniform fluid,
$U^{0}=1,$ $U^{j}=0.$ We assume the equation of state%
\begin{equation}
p_{{m}}=\epsilon_{0}\rho_{{m}}%
\end{equation}
where $\epsilon_{0}=1/3$ for radiation, and $\epsilon_{0}=0$ for classical
matter. The coupling to the scalar field is specified via an interaction
Lagrangian density $\mathcal{L}_{{int}}$. We give some examples of
possible interactions.

The simplest interaction is a direct interaction with a real scalar field:
$\mathcal{L}_{{int}}=-\lambda\rho_{{m}}\phi$. Current-current
interaction with a complex scalar field $(N=2)$ can be constructed as follows.
Represent the scalar field in terms of $\phi=2^{-1/2}\left(  \phi_{1}%
+i\phi_{2}\right)  $ and its complex conjugate $\phi^{\ast}$, or in terms of
the phase representation $\phi=F\exp\left(  i\sigma\right)  $. The conserved
scalar current density in the absence of interaction is $J_{\mu}^{{sc}%
}=\left(  2i\right)  ^{-1}\left(  \phi^{\ast}\partial_{\mu}\phi-\phi
\partial_{\mu}\phi^{\ast}\right)  =F^{2}\partial_{\mu}\sigma$. The current
density of a perfect fluid is $J_{\nu}^{{m}}=\rho_{{m}}U_{\nu}$. The
current-current interaction corresponds to
\begin{eqnarray}
\mathcal{L}_{{int}}  &  =-\lambda g^{\mu\nu}J_{\mu}^{{sc}}J_{\nu
}^{{m}}=\lambda\rho_{{m}}g^{\mu\nu}F^{2}\left(  \partial_{\mu}%
\sigma\right)  U_{\nu}\nonumber\\
&  =-\lambda\rho_{{m}}F^{2}\dot{\sigma}{ \ \ \ \ (spatially uniform
system)}%
\end{eqnarray}

Returning to the general case, we can decompose the total energy-momentum
tensor of scalar field and perfect fluid as follows:%
\begin{equation}
T^{\mu\nu}=T_{{sc}}^{\mu\nu}+T_{{m}}^{\mu\nu}+T_{{int}}^{\mu
\nu}%
\end{equation}
We assume%
\begin{equation}
T_{{int}}^{\mu\nu}=-g^{\mu\nu}\mathcal{L}_{{int}}%
\end{equation}
which leads to an interaction energy density $\rho_{{int}}$ and pressure
$p_{{int}}$:%
\begin{eqnarray}
\rho_{{int}}  &  =-\mathcal{L}_{{int}}\nonumber\\
p_{{int}}  &  =\mathcal{L}_{{int}}%
\end{eqnarray}
The equation of motion for the perfect fluid comes from the conservation law
$T_{;\mu}^{\mu\nu}=0$, which for a spatially uniform system reduces to%
\begin{equation}
\dot{\rho}+3H\left(  \rho+p\right)  =0 \label{fluid}%
\end{equation}
where%
\begin{eqnarray}
\rho &  =\rho_{{sc}}+\rho_{{m}}+\rho_{{int}}=\frac{1}{2}%
\sum_{n}\dot{\phi}_{n}^{2}+V{ }+\rho_{{m}}+\mathcal{L}_{{int}%
}\nonumber\\
p  &  =p_{{sc}}+p_{{m}}+p_{{int}}=\frac{1}{2}\sum_{n}\dot{\phi
}_{n}^{2}-V{ }+\epsilon_{0}\rho_{{m}}-\mathcal{L}_{{int}}
\label{rho}%
\end{eqnarray}

We can rewrite (\ref{fluid}) in a more useful form. The equation of motion for the scalar field is
\begin{equation}
\ddot{\phi}_n=-3 H \dot{\phi}_n - \frac{\partial V}{\partial\phi_n} +\frac{\partial {\cal L}_{{int}}}{\partial\phi_n} 
\end{equation}
Multiply both sides by $\dot{\phi}_{n}$ and summing over $n$, we obtain
\begin{eqnarray}
\frac{1}{2}\frac{d}{dt}\sum_{n}\dot{\phi}_{n}^{2}  &  =-3H\sum_{n}\dot{\phi
}_{n}^{2}-\sum_{n}\frac{\partial V}{\partial\phi_{n}}\dot{\phi}_{n}+\sum
_{n}\frac{\partial\mathcal{L}_{{int}}}{\partial\phi_{n}}\dot{\phi}_{n}%
\end{eqnarray}
We note that%
\begin{equation}
\frac{dV}{dt}=\sum_{n}\frac{\partial V}{\partial\phi_{n}}\dot{\phi}_{n}%
+\frac{\partial V}{\partial\Lambda}\dot{\Lambda}%
\end{equation}
Thus%
\begin{equation}
\sum_{n}\frac{\partial V}{\partial\phi_{n}}\dot{\phi}_{n}=\frac{dV}{dt}%
-\frac{\partial V}{\partial\Lambda}\dot{\Lambda}%
\end{equation}
Using this and the fact that $\Lambda=a^{-1}$, we get%
\begin{equation}
\frac{d}{dt}\left(  \frac{1}{2}\sum_{n}\dot{\phi}_{n}^{2}+V\right)
=-3H\sum_{n}\dot{\phi}_{n}^{2}+\sum_{n}\frac{\partial\mathcal{L}_{{int}}%
}{\partial\phi_{n}}\dot{\phi}_{n}+a\frac{\partial V}{\partial a}H
\end{equation}
Now, using (\ref{rho}), we can rewrite (\ref{fluid}) as%
\begin{equation}
\frac{d}{dt}\left[  \frac{1}{2} \sum_{n}\dot{\phi}_{n}^{2}+V{ }%
+\rho_{{m}}+\mathcal{L}_{{int}}\right]  
=-3H\left[  %
\sum_{n}\dot{\phi}_{n}^{2}+\left(  1+\epsilon_{0}\right)  \rho_{{m}%
}\right]
\end{equation}
Using the equation before this, we finally obtain
\begin{equation}
\frac{d\rho_m}{dt}=-3H\left(  1+\epsilon_{0}\right)  \rho_m  -\sum_{n}\frac{\partial\mathcal{L}_{{int}}}{\partial\phi_{n}}\dot{\phi
}_{n}-\frac{d\mathcal{L}_{{int}}}{dt}-a\frac{\partial V}{\partial a}H
\end{equation}

In summary, the cosmological equations are, with $H=\dot{a}/a$,
\begin{eqnarray}
\dot{H}  &  =\frac{k}{a^{2}}-4\pi G\left[  \sum_{n}\dot{\phi}_{n}^{2}+\left(
1+\epsilon_{0}\right)  \rho_{{m}}\right]  +\frac{1}{3}a\frac{\partial
V}{\partial a}\nonumber\\
\ddot{\phi}_{n}  &  =-3H\dot{\phi}_{n}-\frac{\partial V}{\partial\phi_{n}%
}+\frac{\partial\mathcal{L}_{{int}}}{\partial\phi_{n}}\nonumber\\
\dot{\rho}_{{m}}  &  =-3H\left(  1+\epsilon_{0}\right)  \rho_{{m}%
}-\sum_{n}\frac{\partial\mathcal{L}_{{int}}}{\partial\phi_{n}}\dot{\phi
}_{n}-\frac{d\mathcal{L}_{{int}}}{dt}  -a \frac{\partial V}{\partial a} H  \nonumber\\
H^{2}  &  =\frac{2}{3}\left(  \frac{1}{2}\sum_{n=1}^{N}\dot{\phi}_{n}%
^{2}+V+\rho_{{m}}\right)  -\frac{k}{a^{2}}%
\end{eqnarray}
The last equation is a constraint on initial conditions, and is preserved by
the equations of motion. This defines a self-consistent initial-value problem.

Analytical and numerical studies show that matter creation is inefficient, and
that no decoupling occurs between expansion and matter dynamics. 
This failure motivates the consideration of phase dynamics in a complex field, as describe in the next paper of this series.

\bigskip

\newpage


\end{document}